\documentclass[aps,twocolumn,superscriptaddress]{revtex4-1}

\usepackage[utf8]{inputenc}
\usepackage{hyperref}
\usepackage{graphicx}
\usepackage{amsmath}

\usepackage[usenames,dvipsnames]{color}
\usepackage{cancel}

\begin{document}

\title{Controlling inertial focussing using rotational motion}
\author{Christopher Prohm}
\email[Please direct all communication at: ]{christopher.prohm@tu-berlin.de}
\affiliation{Institute of Theoretical Physics, Technische Universit\"at  Berlin, Hardenbergstr. 36, 10623 Berlin, Germany}
\author{Nikolas Zöller}
\affiliation{Institute of Theoretical Physics, Technische Universit\"at  Berlin, Hardenbergstr. 36, 10623 Berlin, Germany}
\author{Holger Stark}
\affiliation{Institute of Theoretical Physics, Technische Universit\"at  Berlin, Hardenbergstr. 36, 10623 Berlin, Germany}

\begin{abstract}
In inertial microfluidics lift forces cause a particle to migrate across streamlines to specific positions in the
cross section of a microchannel. We control the rotational motion of a particle and demonstrate that this allows to
manipulate the lift-force profile and thereby the particle's equilibrium positions. We perform two-dimensional simulation 
studies using the method of multi-particle collision dynamics. Particles with unconstrained rotational motion 
occupy stable equilibrium positions in both halfs of the channel while the center is unstable. When an external torque is applied to the
particle, two equilibrium positions annihilate by passing a saddle-node bifurcation and only one stable fixpoint remains so that
all particles move to one side of the channel.  In contrast, non-rotating particles accumulate in the center and are pushed into one half 
of the channel when the angular velocity is fixed to a non-zero value.
\end{abstract}

\maketitle

%%%%%%%%%%%%%%%%%%%%%%%%%%%%%%%%%%%%%%%%%%%%%%%%%%%%%%%%%%%%%%%%%%%%%%%%%%%%%%%%%%%%%%%%%%%%%%%%%%
%%%%%%%%%%%%%%%%%%%%%%%%%%%%%%%%%%%%%%%%%%%%%%%%%%%%%%%%%%%%%%%%%%%%%%%%%%%%%%%%%%%%%%%%%%%%%%%%%%
%%                                         Introduction                                         %%
%%%%%%%%%%%%%%%%%%%%%%%%%%%%%%%%%%%%%%%%%%%%%%%%%%%%%%%%%%%%%%%%%%%%%%%%%%%%%%%%%%%%%%%%%%%%%%%%%%
%%%%%%%%%%%%%%%%%%%%%%%%%%%%%%%%%%%%%%%%%%%%%%%%%%%%%%%%%%%%%%%%%%%%%%%%%%%%%%%%%%%%%%%%%%%%%%%%%%
\section{Introduction}

%% General Intro for particle transport inertial microfluidics + application 
The transport of particles in a fluid is a recurring problem on very different length scales.
It ranges from dust particle in oil transport  to cells in biomedical applications.
In all these systems fluid inertia cannot be neglected and inertial migration of particles across
streamlines is observed. 
In particular for biomedical applications a large number of microfluidic devices using inertial migration 
has been proposed in recent years \cite{DiCarlo2009Inertial,Park2009,Mach2011,Hur2011Deformability,Mach2010,Guan2013Spiral,Dudani2013Pinched}.
They are used for cell counting, cell sorting, and mechanical phenotyping. 

%% From control of geometry to external fields
The devices are carefully designed for controlling the motion of the dispersed colloidal particles.
Commonly they use a combination of special shapes for the microchannel cross sections
and the action of inertial lift forces to tailor the equilibrium positions of the particles
at the channel outlet \cite{Guan2013Spiral,Mach2010}.
Examples are regular extrusions along the channel to create microscale vortices \cite{Park2009,Mach2011}, 
a trapezoidal channel cross section to enhance particle separation \cite{Guan2013Spiral},
or the controlled placement of pillars into the fluid flow to 
guide particles through the channel \cite{Amini2013Engineering}.
%
%% Translational control 
Aside from geometry, microfluidic devices commonly employ external control fields both at 
low and modest Reynolds numbers $\mathrm{Re}$.
Sheath flows are used to focus particles \cite{Xuan2010} and to analyze the mechanical properties of cells 
in order to detect cancer \cite{Dudani2013Pinched}. 
Optical tweezers are employed, in combination with feedback control \cite{Applegate2007Optically,Wang2011Enhanced,Munson2010Image} 
or as optical lattices \cite{MacDonald2003}, to enhance particle separation.

%% Rotational control (from optical tweezers to magnets)
This article deals with rotating particles in microfluidic channels.
Using the angular momentum of light, optical tweezers perating in the Laguerre-Gaussian mode can transfer
angular momentum to birefringent or shape anisotropic particles and 
rotate them \cite{Grier2003,Padgett2011}.
% In external magnetic fields 
Superparamagnetic beads align with an intrinsic axis along an external magnetic field
due to a small anisotropy in their magnetic susceptibility \cite{Neuman2007Single}.
In particular, they can follow a rotating magnetic field and assume its angular velocity \cite{Janssen2009Controlled}.
Rotating particles were used  to study the mechanical properties of DNA \cite{Lipfert2010Magnetic}
and show rich patterns of self-organization \cite{Grzybowski2000Dynamic,Grzybowski2002Dynamic}.
Furthermore, rotating colloids were employed to construct micropumps both in experiments \cite{Bleil2006Field,Ladavac2004Microoptomechanical} 
and simulations \cite{Gotze2010Flow,Goetze2011Dynamic}.

%% Inertial microfludicis intro
Segr\'{e} and Silberberg  were the first to explicitly attribute cross-streamline migration at 
intermediate Reynolds number to fluid inertia \cite{Segre1961}.
%
%% Describe previous theoretical works inertial microfluidics 
The theoretical analysis of inertial effects in microfluidics is complicated by the
nonlinear convection term of the Navier-Stokes equations.
Analytic studies \cite{Ho1974,Asmolov1999} are only partly applicable as they assume 
particle sizes much smaller than the channel diameter, which is not valid in many microfluidic devices.
Here numerical simulation studies are important.
Previous investigations used the finite-element method \cite{DiCarlo2009,Feng1994-II}, 
lattice-Boltzmann simulations \cite{Chun2006,Prohm2014Feedback},
and the method of multi-particle collision dynamics \cite{Prohm2012}.

%% Outlook
Inertial lift forces are a common means to describe cross-streamline migration of particles in
inertial microfluidics. In this article we investigate how the controlled rotational motion of a
colloidal particle influences the lift-force profile and thereby the equilibrium positions in the channel
cross section. We will perform two-dimensional simulation studies based on the method of
multi-particle collision dynamics. When the particle's rotation is not constrained, it occupies stable
equilibrium positions in both halfs of the channel while the center is unstable. We first control particle
rotation by an external torque and show that with increasing torque particles only occupy one half
of the channel when passing
a saddle-node bifurcation. In contrast, non-rotating particles accumulate in the center 
and are pushed to one side of the channel for non-zero but fixed angular velocity.

% Outline  
The article is organized as follows. 
In Sect.~\ref{sec:methods} we describe the system geometry and shortly introduce the simulation method.
In Sect.\ \ref{sec:uncontrolled} we review particle motion in the inertial regime without rotation control.
We study the effect on the lift force profile when an external torque is applied in Sect.\ \ref{sec:torque} and 
when the angular velocity is controlled in Sect.~\ref{sec:angular-velocity}. Finally, we close with a summary
and concluding remarks in Sect.\ \ref{sect.concl}.

%%%%%%%%%%%%%%%%%%%%%%%%%%%%%%%%%%%%%%%%%%%%%%%%%%%%%%%%%%%%%%%%%%%%%%%%%%%%%%%%%%%%%%%%%%%%%%%%%%
%%%%%%%%%%%%%%%%%%%%%%%%%%%%%%%%%%%%%%%%%%%%%%%%%%%%%%%%%%%%%%%%%%%%%%%%%%%%%%%%%%%%%%%%%%%%%%%%%%
%%                                         Methods                                              %%
%%%%%%%%%%%%%%%%%%%%%%%%%%%%%%%%%%%%%%%%%%%%%%%%%%%%%%%%%%%%%%%%%%%%%%%%%%%%%%%%%%%%%%%%%%%%%%%%%%
%%%%%%%%%%%%%%%%%%%%%%%%%%%%%%%%%%%%%%%%%%%%%%%%%%%%%%%%%%%%%%%%%%%%%%%%%%%%%%%%%%%%%%%%%%%%%%%%%%
\section{Methods}
\label{sec:methods}

\subsection{System}

\begin{figure}
\includegraphics[width=0.7\columnwidth]{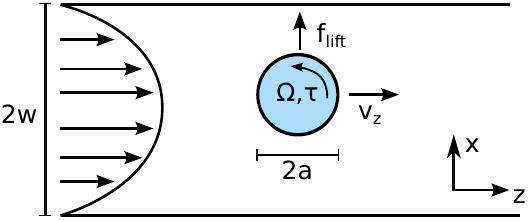}
\caption{ 
Two-dimensional channel geometry with width $2 w$. 
A parabolic Poiseuille flow 
at Reynolds number $\mathrm{Re}$ drives
a circular particle with diameter $2 a$.
The particle moves with an axial velocity $v_z$ and angular velocity $\Omega$. 
It experiences the inertial lift force $f_\mathrm{lift}$ in lateral direction.
}
\label{fig:schematic}
\end{figure}

%% Channel 
We consider a two-dimensional channel geometry with width $2w$ and length $L$ as sketched in Fig.~\ref{fig:schematic}.
The channel is filled by a Newtonian fluid with kinematic viscosity $\nu$ and density $\rho$.
We apply a pressure driven, parabolic Poiseuille flow with maximum flow velocity $u_\mathrm{max}$ at the channel center
and characterize the relevance of fluid inertia by the channel Reynolds number $\mathrm{Re} = 2 u_\mathrm{max} w / \nu$.
The vorticity $\Omega_0(\vec{x}) = [\vec{\nabla} \times \vec{u}(\vec{x}) ]_y / 2$ assumes its maximum value at the 
channel wall and equals $\Omega_\mathrm{max} =  \nu \mathrm{Re}  / 2 w^2$.
In units of the MPCD method introduced below, the respective channel width and length are $2w = 36$ and length $L = 100$.
Along the channel axis we implement periodic boundary conditions
and choose the coordinate system such that the $z$ axis points along the axial direction and the $x$ axis along the lateral direction.

%% Colloid
Inside the channel we place a neutrally buoyant, circular particle with radius $a$.
We denote its lateral distance from the center line by $x$.
The particle moves with an axial velocity $v_z$ along the channel and experiences a lateral lift force $f_\mathrm{lift}$
while rotating with an angular velocity $\Omega$.

\subsection{Multi-particle collision dynamics}
To simulate the flow of a
Newtonian fluid, we use the method of Multi-Particle Collision Dynamics (MPCD) \cite{Malevanets1999,Gompper2009,Kapral2008}.
MPCD is a mesoscopic Navier-Stokes solver which includes thermal fluctuations.
It uses point particles which stream during a time intervall $\Delta t_c$ with their respective velocities. 
Then during the collision step the simulation volume is divided into cells. 
The velocities of the particles in each cell are redistributed by a collision rule such that linear momentum is preserved.
Here we employ a specific variant of MPCD called MPCD-AT+a, which simulates fluid flow at  constant temperature and
also conserves local angular momentum which is especially important when considering rotating surfaces \cite{Goetze2007}. 
Details of our implementation and also how to generate the pressure driven Poiseuille flow are described in our previous
work on inertial microfluidics with three-dimensional flows \cite{Prohm2012}.

%% Details to include for full description:
%%  - Analytical formula for the viscosity
%%  - Basic implementation
%%  - Ghost particles
%%  - Grid shifting 
%%  - Fluid object coupling

%% Parameters 
In our simulations we choose all quantities in MPCD units. In particular, we use
the edge length of the quadratic collision cells $a = 1$, the collision time $\Delta t_c = 0.1$, 
density $\rho = 10$, and temperature $T = 1$.
These parameters result in a kinematic viscosity $\nu = 0.42$ which we calculate from analytic expressions 
derived in \cite{Noguchi2008}.
As already explained in Ref.\ \cite{Prohm2012}, thermal fluctuations are too strong in the simulations.
However, by averaging over a sufficient number of realizations, we are able to determine lift-force profiles which 
we are mainly interested in.

\subsection{Lift-force profiles and probability distributions from MPCD simulations}
To determine lift forces acting on a particle and its axial velocities as a function of particle position, 
we keep the lateral position $x$ fixed by always setting the lateral velocity $v_x$ to zero.
The motion along the channel axis and the rotational motion of the particle are not constrained.
We determine the lift force $f_\mathrm{lift}$ by averaging the momentum transfer from the fluid particles on the
colloidal particle in steady state over the last 49000 time steps.
Averaging over the same time steps, also gives the steady state axial velocity $v_z$ and the angular velocity $\Omega$ of the particle.
%
%% Constrained angular velocity 
In Sect.\ \ref{sec:angular-velocity} we will consider particle motion with constrained angular velocity. 
Then we also calculate the torque $\tau$ exerted by the fluid onto the particle by averaging the angular momentum transfer in steady state.

%% Kernel density estimation 
We determine the center-of-mass distribution function of the colloidal particle by letting it move 
without any constraints through the channel.
From the particle trajectories we calculate the distribution functions using 
a kernel density estimate \cite{BookSilverman} with a bandwidth $b = 0.03w$ as exlained in \cite{Prohm2012}.

%%%%%%%%%%%%%%%%%%%%%%%%%%%%%%%%%%%%%%%%%%%%%%%%%%%%%%%%%%%%%%%%%%%%%%%%%%%%%%%%%%%%%%%%%%%%%%%%%%
%%%%%%%%%%%%%%%%%%%%%%%%%%%%%%%%%%%%%%%%%%%%%%%%%%%%%%%%%%%%%%%%%%%%%%%%%%%%%%%%%%%%%%%%%%%%%%%%%%
%%                                            Results                                           %%
%%%%%%%%%%%%%%%%%%%%%%%%%%%%%%%%%%%%%%%%%%%%%%%%%%%%%%%%%%%%%%%%%%%%%%%%%%%%%%%%%%%%%%%%%%%%%%%%%%
%%%%%%%%%%%%%%%%%%%%%%%%%%%%%%%%%%%%%%%%%%%%%%%%%%%%%%%%%%%%%%%%%%%%%%%%%%%%%%%%%%%%%%%%%%%%%%%%%%

\section{Uncontrolled rotational motion}
\label{sec:uncontrolled}

%% Distributions
\begin{figure}
\includegraphics[width=\columnwidth]{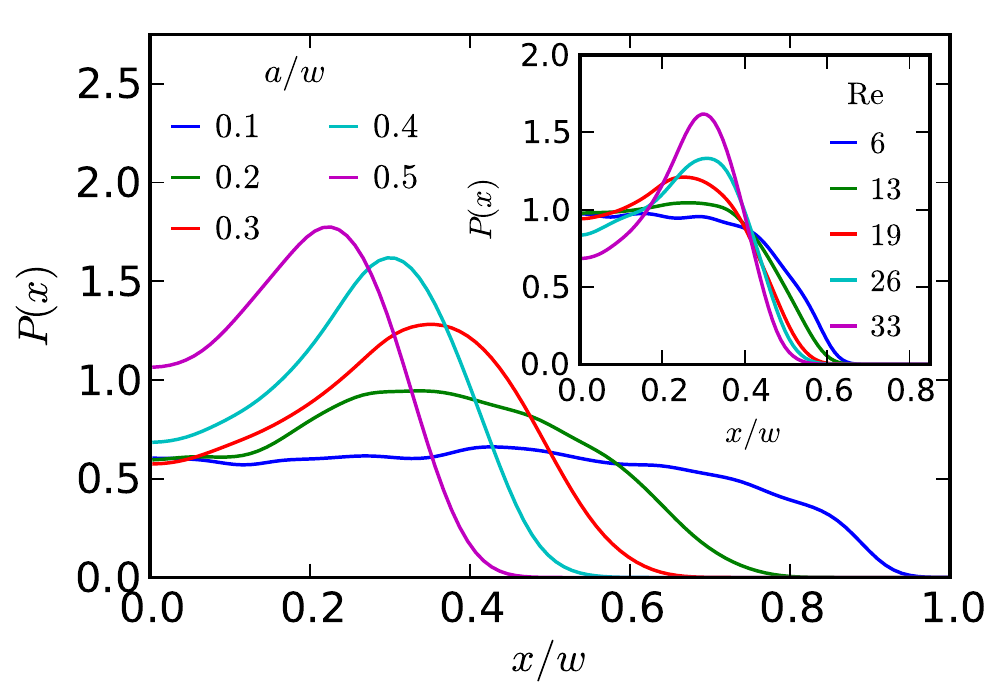}
\caption{
Lateral center-of-mass distribution function $P(x)$ for different particle sizes $a/w$
at Reynolds number $\mathrm{Re} = 33$. 
Rotational particle motion is not controlled.  
Inset: Lateral distribution function for different $\mathrm{Re}$ and $a/w = 0.4$.
}
\label{fig:distribution-free}
\end{figure}

First we illustrate the basic features of particle motion inside a microchannel in the inertial regime
at intermediate Reynolds number when particle rotation is not controlled.
Due to the inertial lift force, the particle migrates across streamlines counterbalanced
bei thermal diffusion. 
As a result, a stationary center-of-mass distribution function $P(x)$ in lateral direction forms, which we plot in Fig.~\ref{fig:distribution-free}.
We observe how $P(x)$ gets narrower and shifts towards the channel center with increasing particle size $a/w$.
The peak is located at the position of zero lift force.
In addition, the distribution function becomes narrower with increasing Reynolds number as demonstrated in the inset.

%% Forces
\begin{figure}
\includegraphics[width=\columnwidth]{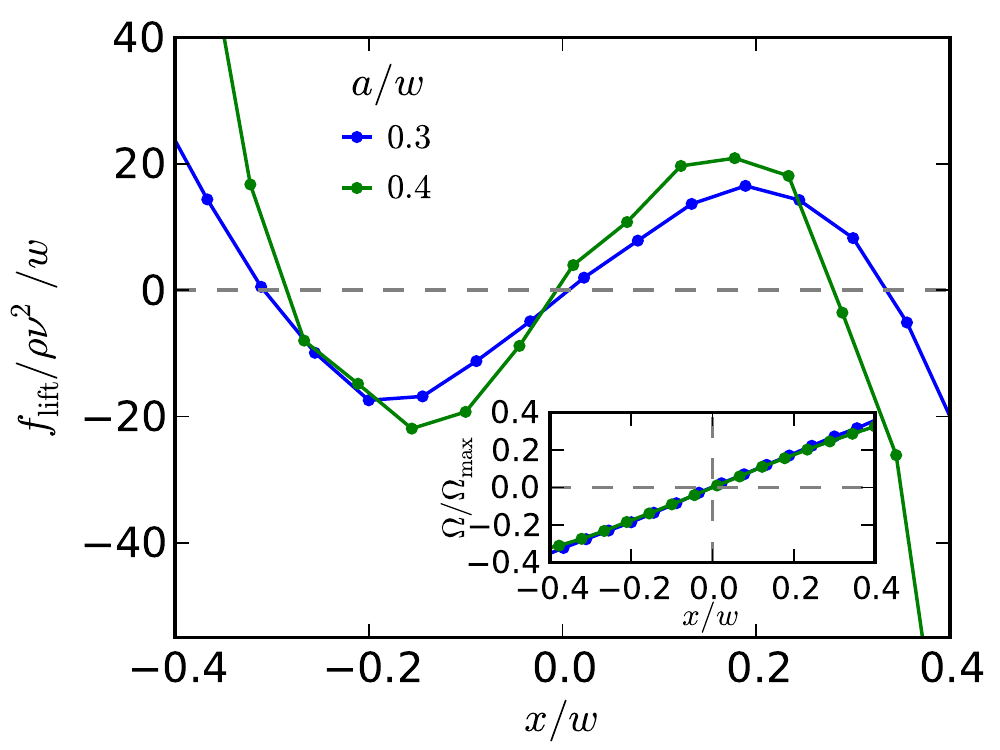}
\caption{
Lift-force profiles $f_\mathrm{lift}$ plotted versus particle position $x$ 
for two particle sizes $a/w = 0.3$ and 0.4 at $\mathrm{Re} = 33$.
The inset shows the angular velocity $\Omega$ of the particle in units of 
the maximum vorticity $\Omega_\mathrm{max}$ plotted versus $x$.
}
\label{fig:free-forces}
\end{figure}

In Fig.\ \ref{fig:free-forces} we plot the lift-force profile for particle sizes $a = 0.3w$ and $0.4w$ 
at channel Reynolds number $\mathrm{Re} = 33$.
The channel center at $x=0$ with $f_{\mathrm{lift}} = 0$ is an unstable fixpoint 
since any disturbance drives the particle towards the channel walls.
However, close to a wall the particle is pushed towards the channel center.
Both effects chancel at about $x \approx \pm 0.3w$ and a stable equilibrium position with $f_{\mathrm{lift}} = 0$ occurs.
Furthermore, we observe that the strength of the lift force increases with increasing particle size.
The particle rotates with an angular velocity $\Omega$ which is linear in the distance from the center 
as the inset of Fig.~\ref{fig:free-forces} shows.
Indeed, from Faxen's law  \cite{Brenner1981Translational}, we expect $\Omega$ to be equal to the local vorticity in an unconstrained fluid, which 
for an unperturbed Poiseuille flow is linear in $x$. 
Small deviations due to hydrodynamic interactions with the bounding walls
will be illustrated in more detail in Fig.\ \ref{fig:torque-angular-velocity}.
Both, the distribution functions and the lift-force profiles show the same qualitative behavior as in simulations of three-dimensional
systems \cite{Prohm2012,DiCarlo2009}.
As demonstrated in \cite{Prohm2012},  we can interpret the lateral distribution function as a Boltzmann distribution since 
the particle performs thermal motion in a potential associated with the inertial lift force.

One can explain the lateral lift force as the result of a difference in dynamic pressure that occurs when in a
shear flow the flow velocities close to the particle surface vary \cite{Matas2004}.
In particular, in this interpretation the inertial lift force is caused by the curvature of the Poiseuille flow  profile. 
The absolute relative velocity between the colloid and fluid is higher on the wall side than on the side facing
the channel centerline. 
This creates a lower dynamic pressure on the wall side and the particle is pushed towards the wall.
Following this explanation, we therefore expect  that changing the angular velocity of the colloid 
alters the lift-force profile since the relative velocities on the two colloid sides change in the opposite way. 
In the following two sections we will explore this effect and show how it can be used to control lateral particle 
migration in the microchannel.

%%%%%%%%%%%%%%%%%%%%%%%%%%%%%%%%%%%%%%%%%%%%%%%%%%%%%%%%%%%%%
%%                                    Control by torque                                         %%
%%%%%%%%%%%%%%%%%%%%%%%%%%%%%%%%%%%%%%%%%%%%%%%%%%%%%%%%%%%%%
\section{Control by torque}
\label{sec:torque}

% Change in distributions 
\begin{figure}
\includegraphics[width=\columnwidth]{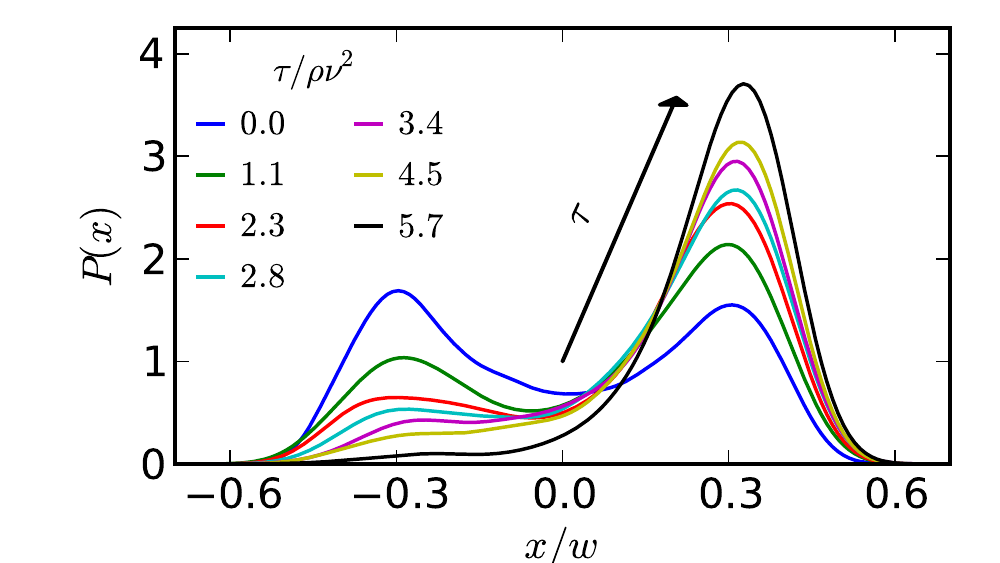}
\caption{Distribution with a additional torque imposed on the particle for 
		 $\mathrm{Re} = 33, a/w = 0.4$. 
		 The torque breaks the symmetry of the system and leads to a preferential 
		 migration in direction of the faster particle side.}
\label{fig:distributions-torque}
\end{figure}

In the present section we investigate how one can control
particle migration by applying a constant torque $\tau$ to the particle.
From Fig.~\ref{fig:distributions-torque}, where we plot the lateral center-of-mass distribution for increasing
$\tau$, we observe that the external torque breaks the original mirror symmetry of the microfluidic system.
With increasing $\tau$ particles preferentially move to the upper half of the channel ($x > 0$) 
and ultimately the asymmetric bimodal distribution assumes a unimodal shape.
The reason becomes clear with the help of Fig.\ \ref{fig:schematic}. 
In the lower half of the channel the particle side facing the wall moves with the flow due to the applied torque which decreases the
difference in relative flow velocities between both sides of the particle. 
In the upper half the difference increases and thereby the difference in dynamic pressure becomes larger. 
This results in a force which pushes the particle against the upper wall.

%% Forces 
\begin{figure}
\includegraphics[width=\columnwidth]{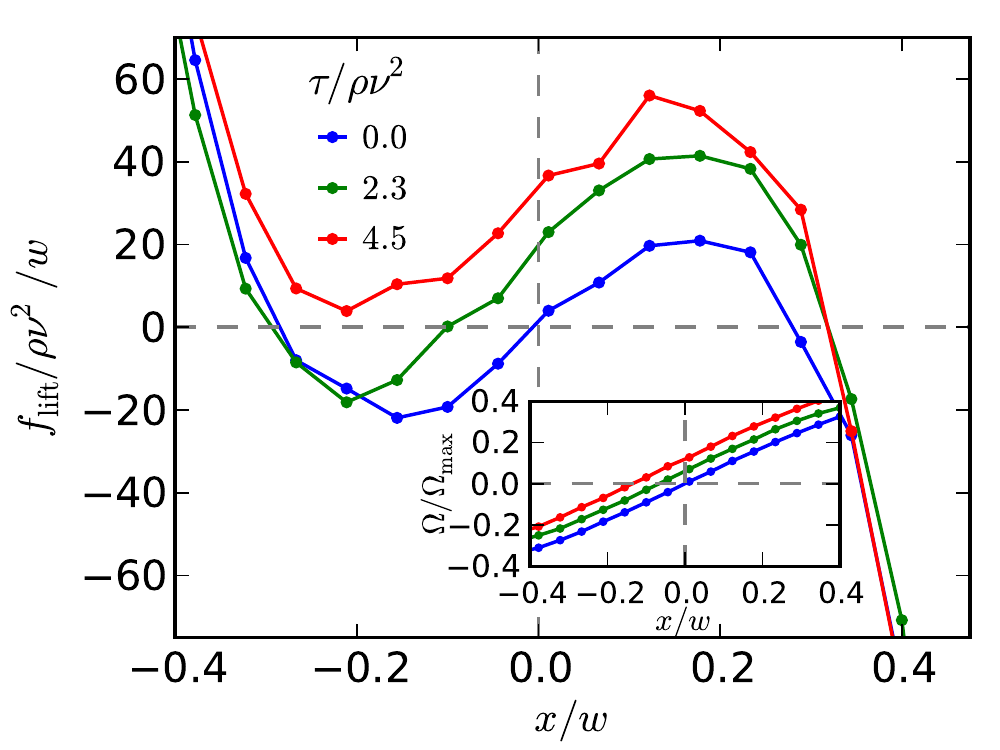}
\caption{Lift-force profile for several applied torque values $\tau$ for $a/w =  0.4$ and at $\mathrm{Re} = 33$. 
Inset: Rotational velocity plotted versus $x$ for the same torque values.
}
\label{fig:force-torque}
\end{figure}

\begin{figure}
\includegraphics[width=\columnwidth]{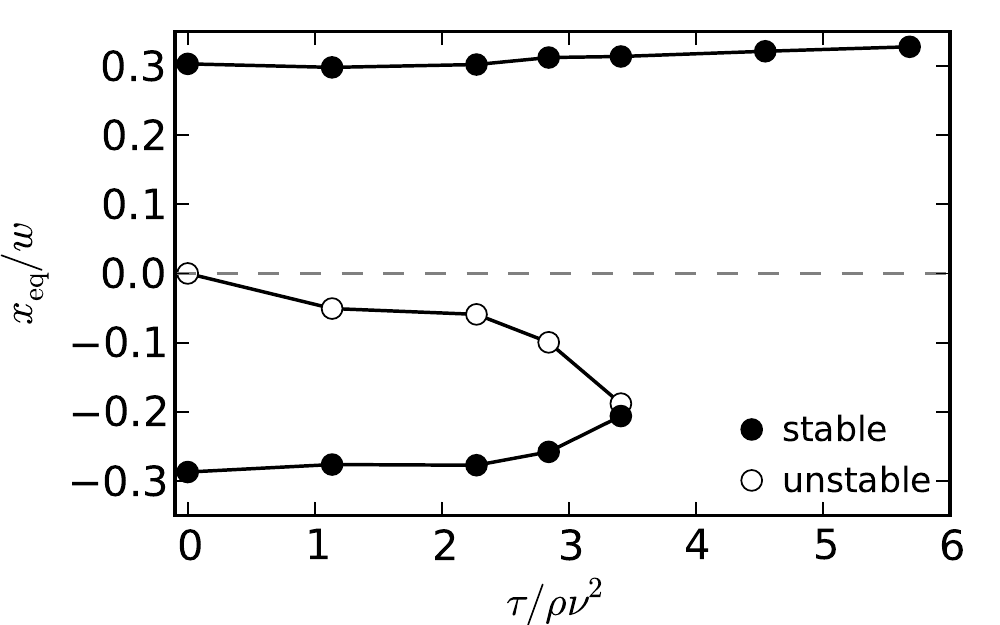}
\caption{
Bifurcation szenario for the stable and unstable equilibrium positions $x_\mathrm{eq}$ plotted as a 
function of the applied torque $\tau$.
The particle size is $a/w = 0.4$ and $\mathrm{Re} = 33$.
A saddle-node bifurcation occurs.
}
\label{fig:bifurcation}
\end{figure}

Modifications in the lift-force profile reflect the change in the distribution function 
as demonstrated in Fig.~\ref{fig:force-torque}.
With increasing torque the whole force profile is shifted upwards. 
Stable and unstable equilibrium positions in the lower half of the channel ($x<0$) move towards each other. 
They ultimately merge (not shown) and vanish completely which corresponds to the case where the distribution function becomes unimodal.
Fig.\ \ref{fig:bifurcation} illustrates the corresponding bifurcation scenario by plotting the equilibrium positions $x_{\mathrm{eq}}$
of the fix points versus the applied torque. 
The stable and unstable fixpoint in the lower half of the channel merge and annihilate each 
other in a saddle-node bifurcation, while the second stable fixpoint slightly moves towards the wall. 
We determined the bifuraction scenario directly from the distribution functions in Fig.\ \ref{fig:distributions-torque}. 
The unstable fix point corresponds to the minimum in the distribution function while the stable equilibrium positions 
follow from the maxima.

% Resulting velocities 
The inset of Fig.~\ref{fig:force-torque} shows how the applied torque $\tau$ 
changes the angular velocity $\Omega$ along the channel cross section.
This is well described by Fax\'en's law \cite{Brenner1981Translational}
\begin{equation}
\Omega - \Omega_0 = \xi_r^{-1} \tau .
\label{eq.faxen}
\end{equation}
However, small deviations occur due hydrodynamic interactions with the channel walls,
as we will demonstrate in the following section and Fig.\ \ref{fig:torque-angular-velocity}.
Here $\Omega_0(x)$ is the vorticity of the unperturbed Poiseuille flow and 
$\xi_r = 4 \pi \rho \nu a^2$
the rotational friction coefficient of a circular disk \cite{Chwang1974Hydromechanics}.
In contrast to an equivalent investigation at small $\mathrm{Re} \approx 0$ \cite{Goetze2010Flow},
we do not observe that the axial particle velocity $v_z$ varies for different torques and compared to the free particle motion (not shown). 
We expect this since the external driving by the Poiseuille flow is quite strong.

%% Size effects
\begin{figure}
\includegraphics[width=\columnwidth]{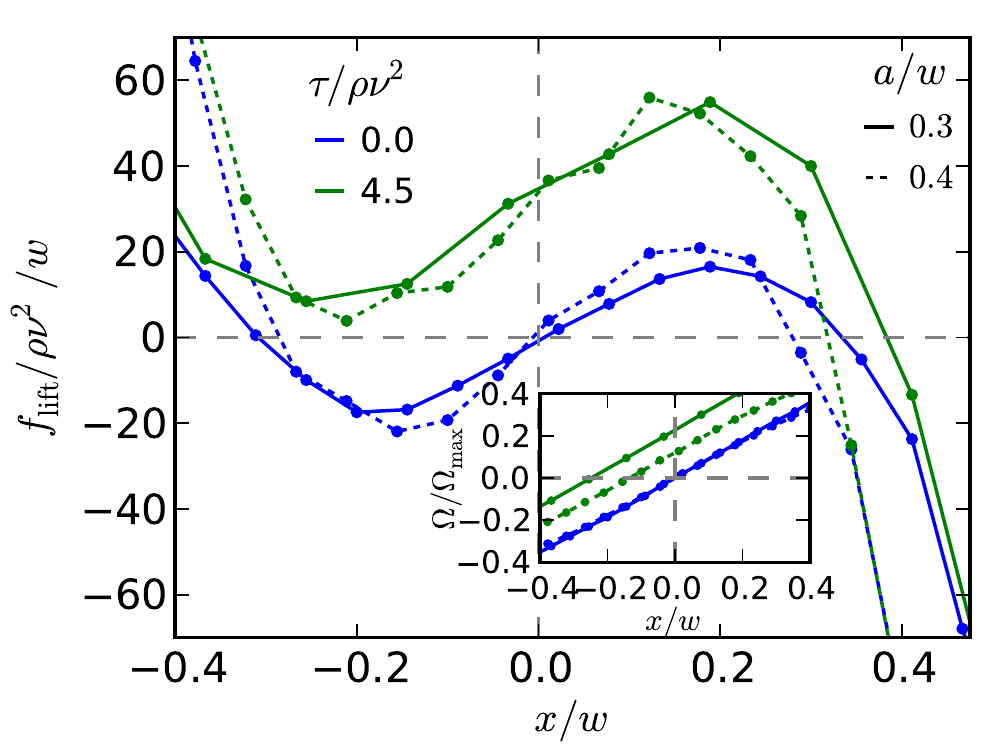}
\caption{Lift-force profiles at $\mathrm{Re} = 33$ for zero torque (blue) and 
$\tau = 0.12 \xi_r \dot{\gamma}_\mathrm{max}$ (green) for two particles sizes $a = 0.3w$ (solid)
and $a = 0.4 w$ (dashed). Inset: Rotational velocity plotted versus $x$ for the same parameters.
}
\label{fig:size-comparison}
\end{figure}

In Fig.\ \ref{fig:size-comparison} we plot the lift-force profile  
for zero torque and $\tau = 4.5 \rho \nu^2$ 
for two particle sizes $a = 0.3w$ and $0.4w$.
Surprisingly, the applied torque shifts the force profiles for both particle sizes upwards by the same force value.
Only close to the wall deviations occur since the torque-induced lift force has to compete with the strong wall-induced lift forces.
Since the rotational friction coefficient of the small particle is
smaller by a factor $(4/3)^2 = 1.77$ compared to the larger particle, we 
observe a stronger increase in the angular velocity as documented by the inset
in Fig.~\ref{fig:size-comparison}. 
We obtain quantitative agreement with Fax\'en's law (\ref{eq.faxen}) with
small deviations when the particle approaches the walls.

%% Demonstrate inertial effect
\begin{figure}
\includegraphics[width=\columnwidth]{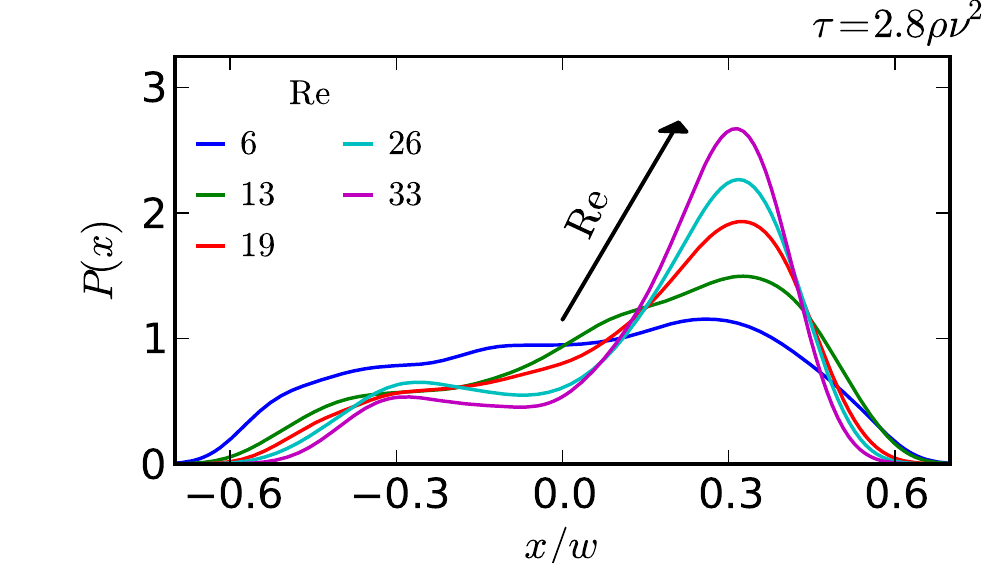}
\caption{
Distribution functions for a fixed torque value $\mathrm{\tau} = 2.8\rho\nu^2$ 
at different Reynolds numbers.
}
\label{fig:distribution-torque-reynolds}
\end{figure}

The contribution to the lift force generated by the applied torque is an inertial force. 
In Fig.~\ref{fig:distribution-torque-reynolds}, we plot
distribution functions for different Reynolds numbers $\mathrm{Re}$ all determined with the same applied torque.
We observe that with decreasing $\mathrm{Re}$ the particle focussing vanishes although the
imposed torque stays the same.

%%%%%%%%%%%%%%%%%%%%%%%%%%%%%%%%%%%%%%%%%%%%%%%%%%%%%%%%%%%%%%%%%%%%%%%%%%%%%%%%%%%%%%%%%%%%%%%%%%
%%                             Control by angular velocity                                      %%
%%%%%%%%%%%%%%%%%%%%%%%%%%%%%%%%%%%%%%%%%%%%%%%%%%%%%%%%%%%%%%%%%%%%%%%%%%%%%%%%%%%%%%%%%%%%%%%%%%
\section{Control by angular velocity}
\label{sec:angular-velocity} 

%% Magnetic field
So far, we controlled the external torque acting on the particles. It adds a constant rotational velocity
to the position dependent part $\Omega_0$, which is given by the local vorticity of the Poiseuille flow. 
Now, we control the particles' rotational velocity $\Omega$ and keep it constant in the channel cross section. 
This will offer new insights.
To impose $\Omega$, we think, for example, about superparamagnetic colloids.
The magnetic susceptibility is typically anisoptropic and the particles align  
with their intrinsic axes along an external field \cite{Neuman2007Single}.
The particle axis can also follow a rotating magnetic field and revolves with the same angular velocity
\cite{Janssen2009Controlled}.

%% Observed distributions for constrained angular velocity 
\begin{figure}
\includegraphics[width=\columnwidth]{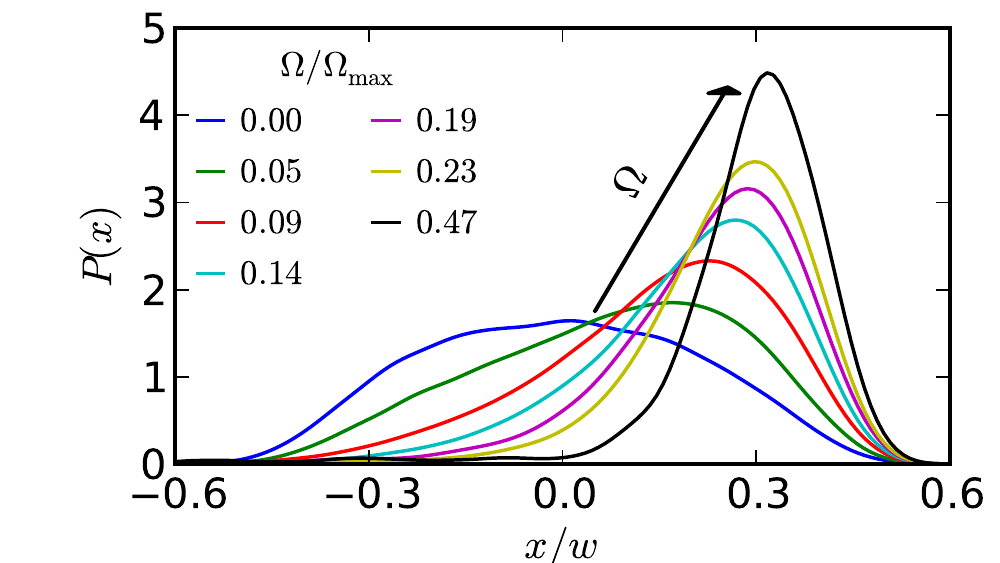}
\caption{
%Lateral 
Particle distribution functions along the lateral direction for different angular velocities $\Omega$
         imposed on a particle with radius $a = 0.4w$ at Reynolds number $\mathrm{Re} = 33$.
        }
\label{fig:distribution-angular-velocity}.
\end{figure}

\begin{figure}
\includegraphics[width=\columnwidth]{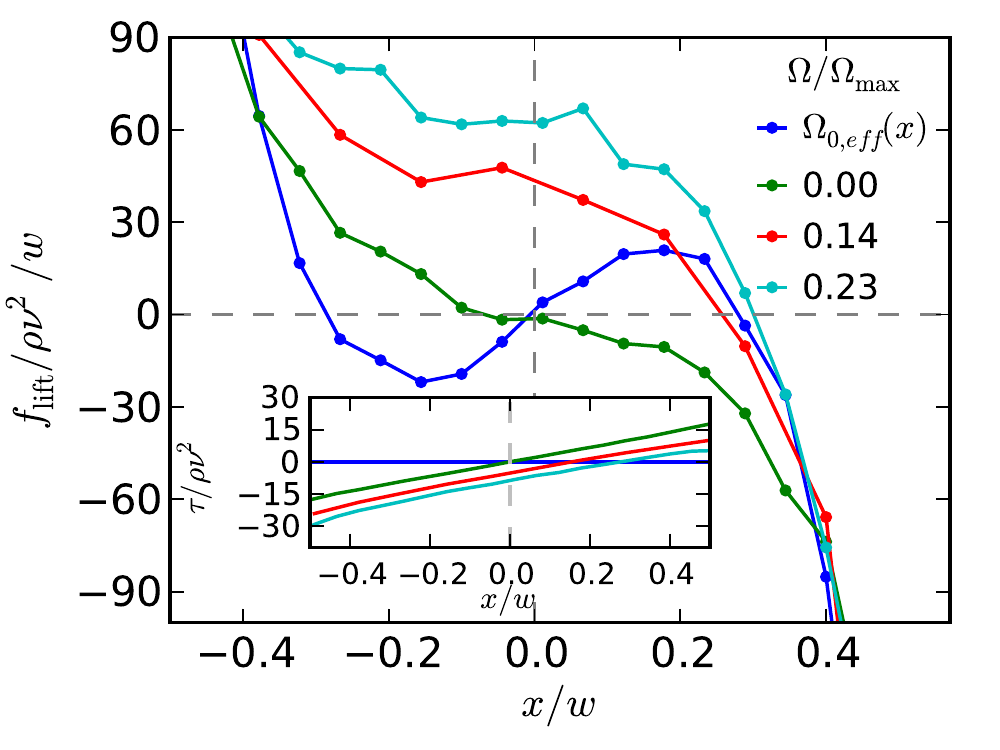}
\caption{
         Lift-force profiles for several imposed angular velocities $\Omega$ and the
         unconstrained motion with $\Omega = \Omega_{0,\mathrm{eff}}(x)$. 
         Inset: Torque necessary to impose a given angular velocity at different lateral positions $x$.
}
\label{fig:force-angular-velocity}
\end{figure}

Figure\ \ref{fig:distribution-angular-velocity} shows particle distribution functions along the lateral direction 
for different angular velocities $\Omega$.
The particle has a radius $a = 0.4w$ and the Reynolds number is $\mathrm{Re} = 33$. 
We observe that the distribution $P(x)$ is unimodal for all angular velocities in contrast to the bimodal distribution
when the particle rotation is not controlled.
At $\Omega=0$ the particle predominantly stays in the center of the channel.
With increasing angular velocity the maximum of $P(x)$ shifts towards the wall,
in agreement with our findings for the lift-force profile discussed below and illustrated in
Fig. \ref{fig:force-angular-velocity}.
These results show that particle rotation is important for inertial focussing. 
In contrast to findings in \cite{DiCarlo2009}, we are even able to focus the particle to the channel center.
Finally, we note that a negative $\Omega$ shifts the particle into the lower half of the channel. 
So within inertial microfluidics on can position particles inside the channel by controlling their angular velocity.

%% Shift in equilibrium position 
\begin{figure}
\includegraphics[width=\columnwidth]{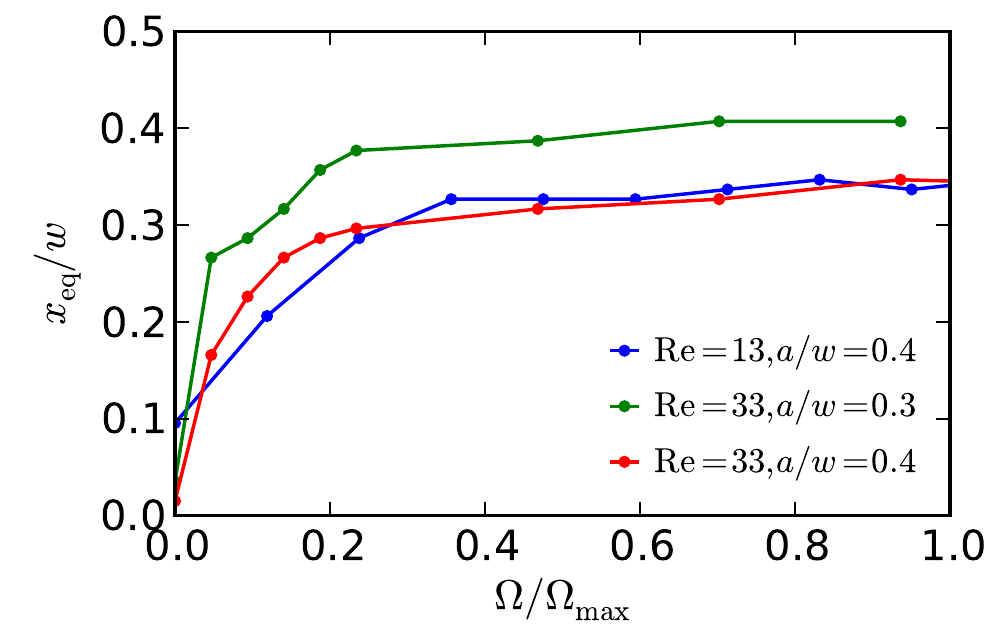}
\caption{
		 Equilibrium position $x_\mathrm{eq}$ as a function of the imposed 
		 angular velocity $\Omega$ for different particle sizes and $\mathrm{Re}$.
		 }
\label{fig:change-xeq-angular-velocity}
\end{figure}

To be more quantitative, we determine the equilibrium position $x_\mathrm{eq}$ from the maximum of the distribution 
function and in Fig.\ \ref{fig:change-xeq-angular-velocity} plot it versus the  angular velocity $\Omega$
for different particle sizes and Reynolds numbers.
%
% Compare size 
Starting from the center for zero angular velocity, the particle position shifts closer to the wall with increasing 
$\Omega$, as already stated.
Similar to the unconstrained case, the equilibrium position of the  smaller particle is always closer to the wall
as the green ($a=0.3w$) and the red ($a=0.4w$) graph illustrate.
%
% Compare Reynolds number 
We also plot the equilibrium position of a particle with radius $a/w = 0.4$ at a lower Reynolds number $\mathrm{Re} = 13$.
Since we rescale $\Omega$ by the maximum vorticity $\Omega_{\mathrm{max}} \propto \mathrm{Re}$,
the absolute values of the imposed angular velocities are only 40\% of the ones at the larger Reynolds number.
Nevertheless, with this scaling we observe a similar behavior of the equilibrium positions for both Reynolds numbers.

%% Forces / Torques needed for angular velocity 

Figure\ \ref{fig:force-angular-velocity} shows how the inertial lift-force profile is 
influenced by the imposed constant angular velocity $\Omega$. 
For comparison, we also show the lift-force profile of an unconstrained particle which has an angular velocity of 
$\Omega_{0,\mathrm{eff}}(x)$ almost equal to the vorticity of the Poiseuille flow $\Omega_0$, as mentiond earlier. 
When we prevent the particle from rotating, it experiences a force which always drives it back to the channel center. 
With increasing angular velocity, the equilibrium position shifts towards the wall 
in agreement with our findings for the distribution function.
Close to the wall the wall lift force dominates and all curves fall on top of each other.
We observe similar behavior for other particle sizes and Reynolds numbers (not shown).
The inset shows the external torque necessary to impose the constant $\Omega$. 
Clearly, for $\Omega =0$, the torque has to compensate the rotational motion induced by the vorticity $\Omega_0$
of the Poiseuille flow $\tau = \xi_r \Omega_0$. 
The straight line is then shifted upwards by $\Delta \tau = \xi_r \Omega$.

%% Torque / angular velocity connection 
\begin{figure}
\includegraphics[width=\columnwidth]{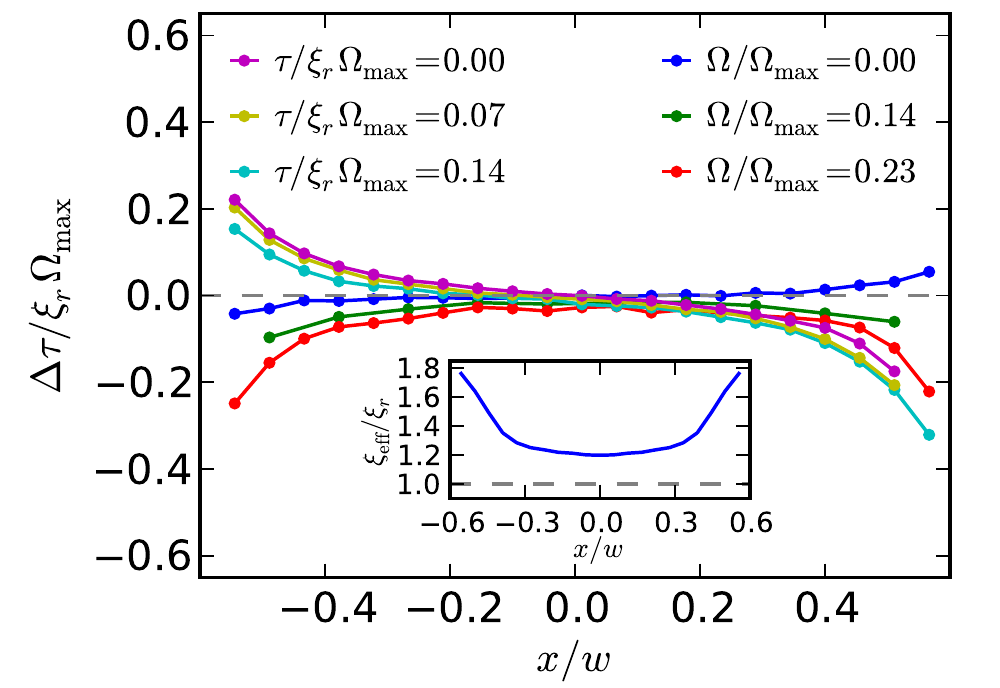}
\caption{
%The deviations 
The deviation $\Delta \tau = \tau - \xi_r(\Omega-\Omega_0)$ from the torque expected according to Fax\'en's law, 
plotted either for prescribed torque $\tau$ or angular velocity $\Omega$ for $a/w = 0.4, \mathrm{Re} = 33$.
Inset: The effective rotational friction coeffcient $ \xi_{\mathrm{eff}} = \tau / (\Omega - \Omega_{0,\mathrm{eff}})$
plotted versus particle position.
}
\label{fig:torque-angular-velocity}
\end{figure}

In the end we test the validity of Fax\'en's law as formulated in Eq.\ (\ref{eq.faxen}) with the vorticity
of the unperturbed Poiseuille flow. Through hydrodynamic interactions with the channel walls
additional flow fields are initiated, which contribute to the flow vorticity. We thereby expect
deviations from Fax\'en's law, when the particle comes close to the wall. To illustrate this,
we plot in Fig.\ \ref{fig:torque-angular-velocity} the deviation from Fax\'en's law, $\Delta \tau = \tau - \xi_r (\Omega-\Omega_0)$,
along the $x$ direction. We either fix particle torque $\tau$ and determine particle rotation $\Omega$ by
Fax\'en's law in Eq.\ (\ref{eq.faxen}) or vice versa. 
Clear deviations, when the particle approaches the walls, are visible.
The deviations are due to two effects. 
The real rotational velocity $\Omega_{0,\mathrm{eff}}$ of the unconstrained
particle deviates from $\Omega_{0}$ and the particle experiences an increased rotational friction, in particular,
close to the channel walls. We determine an effective friction coefficient by averaging 
$ \xi_{\mathrm{eff}} = \tau / (\Omega - \Omega_{0,\mathrm{eff}})$ over all data sets and plot it in the inset of 
Fig.~\ref{fig:torque-angular-velocity}. We observe a pronounced increase as the particle approaches the wall.

%%%%%%%%%%%%%%%%%%%%%%%%%%%%%%%%%%%%%%%%%%%%%%%%%%%%%%%%%%%%%%%%%%%%%%%%%%%%%%%%%%%%%%%%%%%%%%%%%%
%%%%%%%%%%%%%%%%%%%%%%%%%%%%%%%%%%%%%%%%%%%%%%%%%%%%%%%%%%%%%%%%%%%%%%%%%%%%%%%%%%%%%%%%%%%%%%%%%%
%%                                            Conclusion                                        %%
%%%%%%%%%%%%%%%%%%%%%%%%%%%%%%%%%%%%%%%%%%%%%%%%%%%%%%%%%%%%%%%%%%%%%%%%%%%%%%%%%%%%%%%%%%%%%%%%%%
%%%%%%%%%%%%%%%%%%%%%%%%%%%%%%%%%%%%%%%%%%%%%%%%%%%%%%%%%%%%%%%%%%%%%%%%%%%%%%%%%%%%%%%%%%%%%%%%%% 

\section{Conclusion}
\label{sect.concl}

In this article  we investigated the motion of a colloidal particle in a two-dimensional channel geometry 
subject to a pressure driven Poiseuille flow at intermediate Reynolds numbers. We performed a simulation
study using multi-particle collision dynamics. We focussed on how control of the particle's rotation
modifies the lift-force profile and thereby the particle's equilibrium positions in the channel cross section.
For uncontrolled particle motion we observed a strong influence of particle size on the equilibrium position
in accordance with previous results in three dimensions \cite{DiCarlo2009,Prohm2013}.
We also discussed the origin of the lift force. In a parabolic flow the fluid velocities in the frame of the colloid
differ for the two particle sides facing either the wall or the channel center which creates a dynamic pressure
difference \cite{Matas2004}. Since the rotational velocity of the particle directly influences the fluid flow,
we argued that a control of particle rotation also modies the lift-force profile.

To test this prediction, we first applied a constant torque to the particle. 
This breaks the mirror symmetry with respect to the channel axis and the particle
pre\-ferentially migrates to one channel half. With increasing torque, the lift-force profile is shifted 
upwards to larger values meaning that the positions of the unstable and stable fixpoint in the lower
channel half move towards each other until they annihilate in a saddle-node bifurcation and
only the stable fixpoint in the upper channel half remains. Second, we directly controlled the particle's
angular velocity which in experiments can be achieved with a rotating magnetic field. When the particle
does not rotate, it stays in the channel center, and with non-zero rotational velocity it moves into one half of 
the channel. This offers an interesting possibility to effectively separate non-magnetic and magnetic
particles from each other by controlling the rotational velocity of the latter particle type. The response of the
particle to an applied torque or a controlled rotational velocity agrees well with Fax\'en's law. Deviations
occur due to the bounding walls which modify the unconstrained rotation of the particle and its rotational
friction coefficient.

%% Outlook 
% 3d simulations
After having explored the main effect of controlled particle rotation, we will extend our investigations to three-dimensional 
channel geometries. Here, more stable and unstable fixpoints occur\ \cite{Prohm2014Feedback} and it will be
interesting to check how they are modified under controlled rotation. Furthermore, at low Reynolds number 
dense suspensions of colloids subject to a constant torque show very interesting collective behavior \cite{Goetze2010Flow}.
When we combine this with the spontaneous ordering of particles into one-dimensional crystals 
at moderate Reynolds number due to the presence of inertial lift forces \cite{Lee2010,Humphry2010Axial}, we expect a very rich 
and interesting collective behavior, which we plan to explore.

\begin{acknowledgments}
We acknowledge financial support by the Deutsche Forschungsgemeinschaft in the framework of the 
collaborative research center SFB 910.
\end{acknowledgments}

\bibliography{paper}

\end{document}